\documentclass{article}
\usepackage{amssymb}

\input{tcilatex}
\begin{document}

\title{A conjecture about the general cosmological solution of Einstein's
equations}
\author{John D. Barrow \\
DAMTP, Centre for Mathematical Sciences,\\
University of Cambridge,\\
Wilberforce Road, Cambridge CB3 0WA\\
United Kingdom\ \ \ \ \ \ \ \ }
\maketitle
\date{}

\begin{abstract}
We introduce consideration of a new factor, synchronisation of spacetime
Mixmaster oscillations, that may play a simplifying role in understanding
the nature of the general inhomogeneous cosmological solution to Einstein's
equations. We conjecture that, on approach to a singularity, the interaction
of spacetime Mixmaster oscillations in different regions of an inhomogeneous
universe can produce a synchronisation of these oscillations through a
coupling to their mean field in the way first demonstrated by the Kuramoto
coupled oscillator model.
\end{abstract}

\section{ \protect\bigskip Introduction}

The search for an understanding of parts of the general inhomogeneous
solution of the Einstein equations on approach to a cosmological singularity
has been strongly influenced by the work of Belinskii, Khalatnikov and
Lifshitz (BKL) \cite{BKL} on the evolution of the spatially homogeneous
Bianchi type IX ('Mixmaster') universe, first introduced by Misner \cite{mis}%
. The Einstein equations for this cosmological model are not integrable but
clearly exhibit chaotic behaviour in vacuum and with perfect fluids whose
pressure and density obeys $p<\rho $, \cite{JB, JBrep} The smooth
non-separable invariant measure for its complete discrete dynamics has been
solved exactly by Chernoff and Barrow \cite{CB} and simplifies to a double
continued-fraction shift map \cite{JB1, elk}. The general behaviour of the
Mixmaster model within the framework of spatially homogeneous cosmological
solutions of Einstein's equations has led to claims that a general
inhomogeneous solution looks like the spatially homogeneous type IX at the
leading order in a series approximation, or in which the defining constants
(four in vacuum) become independent slowly varying arbitrary functions of
the three spatial variables. This proposal is not as straightforward as it
sounds. We know that compact vacuum and radiation-dominated universes with
Killing vectors, like Bianchi IX, are not linearisation stable. This means
that small perturbations around an exact spatially homogeneous solution of
the defining equations are open dense in first-order series expansions that
do not form the leading term of any series that converges to a true solution
of those equations. This phenomenon is familiar in nonlinear dynamics \cite%
{brill, altas}. As discussed in ref. \cite{BT}, Marsden, Fischer, Moncrief,
and Arms \cite{MA, Arms, FMM, mon} have proved that compact vacuum solutions
of Einstein's equations with Killing symmetries have this subtle property,
which is manifested in various other non-linear systems. In fact, in the
four-function space of the general cosmological solution, small open
neighbourhoods around the homogeneous type IX solution will be dense in
spurious linearisations that are not approximations to a true inhomogeneous
solution. The reason for this is that in the four-function space spanning
cosmological vacuum solutions the points with Killing vectors are conical
and so there are an infinite number of tangents that can be drawn through
the conical point that represents the spatially homogeneous solution. Only
those tangents that run down the sides of the cone correspond to
linearisations of true solutions: the others form a dense set of spurious
linearisations. The non-spurious perturbations must satisfy the Taub
constraint \cite{taub} to ensure they lie down the sides of the cone. This
situation has been examined in detail in the context of perturbations of the
Einstein static universe in refs. \cite{hwang} and \cite{unr}, although the
latter paper makes no reference to linearisation instability.

This situation alerts us to the possibility that the behaviour of the
inhomogeneous general solution might not just look like slightly different
type IX models from place to place. Since the behaviour of the type IX model
is formally chaotic, it is hard to imagine how different locally chaotic
regions are stitched together and remain so as the local oscillations in
each get erratically out of synchronisation. Other investigators have also
drawn attention to additional unusual features, even in the homogeneous
models, like 'spikes' \cite{spike, heinzle}\ caused by steep spatial
gradients in other simpler inhomogeneous \ cosmologies, that were not part
of the first BKL models. However, although these spikes do arise in many
partial differential equations, they might not be generic \cite{measure}.

Motivated by this situation, we propose one new phenomenon that might play a
part in the general solution near Bianchi type IX and alleviate the
'stitching problem' of joining different chaotic oscillatory regions,. This
is described in section 2 and is followed by concluding discussion and
suggestions for further investigations in section 3.

\section{Synchronisation}

The scenario suggested by BKL and adopted by other investigators is that on
approach to the singularity different subregions of space will behave like
separate type IX universes \cite{uggla}. Let us imagine what might happen to
one of those oscillating subregions on approach to the singularity. In
vacuum, it will experience the collective oscillatory gravitational-wave
perturbations of its neighbouring regions. These perturbations might be
expected to be out of phase and effectively random, but the strength of
their effects on our single specimen subregion will grow in strength as the
singularity is approached and each subregion responds to the mean field
created by the oscillations of other regions.

\ This situation is familiar in many areas of science and has been well
described by the famous Kuramoto model of synchronised oscillators \cite%
{kur, kur2, kur3}. The separate oscillators can become synchronised if the
strength of coupling between different oscillators exceeds a critical value.
They become synchronised because each responds to the mean field created by
the oscillations. A simple familiar example is the hand-clapping of an
audience. Clapping starts randomly, but if it strong enough, then soon
everyone seems to be clapping in unison. A plethora of such examples are
known, especially in the biological world. We propose that the same
phenomenon occurs in an inhomogeneous general cosmological solution.
Although different regions might seem like separate type IX universes, the
coupling of their Mixmaster oscillations enables them all to respond to the
mean field and the oscillations should become synchronised. The critical
coupling strength will inevitably be reached on approach to the singularity.
Recall that on any open interval of Mixmaster evolution around the
singularity at $t=0$ there are an infinite number of spacetime oscillations
and they occur far faster than\ the rate of evolution of the mean volume (we
ignore quantum gravitational effects).

The Bianchi IX vacuum equations in standard Hubble-normalised variables are 
\cite{EMac, WainHsu, Ring}, $\ $%
\begin{eqnarray}
N_{1}^{\prime } &=&(q-4\Sigma _{+})N_{1},  \label{1} \\
N_{2}^{\prime } &=&(q+2\Sigma _{+}+2\sqrt{3}\Sigma _{-})N_{2},  \label{2} \\
N_{3}^{\prime } &=&(q+2\Sigma _{+}-2\sqrt{3}\Sigma _{-})N_{3},  \label{3} \\
\Sigma _{+}^{\prime } &=&-(2-q)\Sigma _{+}-3S_{+},  \label{4} \\
\Sigma _{-}^{\prime } &=&-(2-q)\Sigma _{-}-3S_{-},  \label{5}
\end{eqnarray}%
where,

\begin{eqnarray}
q &\equiv &2(\Sigma _{+}^{2}+\Sigma _{-}^{2}),  \label{6} \\
S_{+} &\equiv &\frac{1}{2}\left[ (N_{2}-N_{3})^{2}-N_{1}(2N_{1}-N_{2}-N_{3})%
\right] ,  \label{7} \\
S_{-} &\equiv &\frac{\sqrt{3}}{2}\left[ (N_{3}-N_{2})(N_{1}-N_{2}-N_{3})%
\right] ,  \label{8}
\end{eqnarray}

with the constraint (the generalised vacuum Friedmann equation)

\begin{equation}
\Sigma _{+}^{2}+\Sigma _{-}^{2}+\frac{3}{4}\left[
N_{1}^{2}+N_{2}^{2}+N_{3}^{2}-2(N_{1}N_{2}+N_{2}N_{3}+N_{1}N_{3})\right] =1.
\label{9}
\end{equation}

Here, $\Sigma _{+}(\tau )$ and $\Sigma _{-}(\tau )\ $are the dimensionless
shear variables, $N_{1}(\tau ),N_{2}(\tau ),$ and $N_{3}(\tau )$ define the
Bianchi group structure and the anisotropic 3-curvature components. In
Bianchi IX, the 3-curvature can change sign and is only positive when the
dynamics are close to isotropy. The $\prime $ denotes differentiation with
respect to a time coordinate $\tau $, which is related to the comoving
proper time by $dt/d\tau =1/H,$where $H$ is the mean Hubble expansion rate.
Typically, $H\simeq 1/3t$ as Bianchi IX approaches the initial singularity
at $t=0,$but the ratios of the three expansion scale factors tend to
infinity. Since $\tau \simeq \frac{1}{3}\ln (t)$, the initial singularity
lies at $\tau =-\infty $. In the Friedmann universes, $\Sigma _{+}^{\
}=\Sigma _{-}^{\ }=N_{1}^{\ }=N_{2}^{\ }=N_{3}^{\ }=0$. However, note that
this is not an exact solution of the constraint equation, (\ref{9}), because
there is no closed vacuum Friedmann universe. An axisymmetric, non-chaotic
solution exists with $N_{2}=N_{3}$ and $S_{-}=0$, but this is not of
interest for our discussion as it is non-oscillatory.

Using the constraint equation we have a 4-dimensional set of autonomous
time-dependent ordinary differential oscillator equations of the form,

\begin{equation}
x_{i}^{\prime }=f(x_{i}),i=1,..4.  \label{10}
\end{equation}

This simple form allows us to conjecture what the effects of interactions
between different varieties of type IX dynamics in different places might be
(in effect, allowing the $\Sigma _{+}^{\ },\Sigma _{-}^{\ }$ and $N_{1}^{\
},N_{2}^{\ },N_{3}^{\ }$ to be both space and time variables).

We want to consider the effects of neighbouring oscillatory regions on one
particular subregion as the singularity is approached. The Kuramoto model
imagines that $N$ oscillatory cycles are interacting with a coupling
strength that is the same for each pair of oscillatory regions. It creates
the simplest possible setting for this problem and has turned out to have
unexpectedly wide applications. The oscillators have natural frequencies, $%
\omega _{j}\epsilon (-\infty ,\infty ),$ and their phases are $\theta
_{j}\epsilon \lbrack 0,2\pi ]$. They are assumed to be coupled by the phase
differences of the oscillators in the following simple way

\begin{equation}
\frac{d\theta _{j}}{d\tau }=\omega _{j}+\frac{K}{N}\dsum\limits_{k=1}^{N}%
\sin \{\theta _{k}(\tau )-\theta _{j}(\tau )\},  \label{kur}
\end{equation}%
where the coupling constant is $K\geqslant 0.$ The coherence of the phases,
and the mean field created by the coupled oscillators, is conveniently
measured by the complex order parameter, defined by,

\begin{equation}
r\exp [i\psi ]\equiv \frac{1}{N}\dsum\limits_{j=1}^{N}\exp [i\theta
_{j}(\tau )],  \label{r}
\end{equation}%
where $\left\vert r\right\vert $ measures the degree of coherence: perfect
synchronisation occurs when $\left\vert r\right\vert =1$ and perfect
incoherence, with the $\theta _{j}$ uniformly distributed on $[0,2\pi ),$
occurs when $\left\vert r\right\vert =0;$ $\psi $ is the average phase.

In terms of $r$, the eqns. (\ref{kur}) now become

\begin{equation}
\frac{d\theta _{j}}{d\tau }=\omega _{j}+K\func{Im}\left( r(\tau )\exp
[i\theta _{j}(\tau )]\right) .  \label{new}
\end{equation}
\ 

This shows how the instantaneous mean field of all the oscillators leads to
the evolution of the phases of the oscillators. The crucial effect of the
phase couplings is that as $K$ is allowed to increase a critical transition
occurs when $K=K_{cr\text{ }}$: when $K>K_{cr\text{ }}$the oscillator
frequencies start to become synchronised by virtue of their common responses
to the perturbations by the mean field, although their phases can be
different. Kuramoto showed that

\begin{equation}
r=\sqrt[\ ]{1-\frac{K_{c}}{K},}  \label{11}
\end{equation}

for all $K>K_{cr}$, \cite{strog}.

We suggest that a similar effect occurs in an inhomogeneous generalisation
of the type IX universe, if it exists. The effects of many separate local
regions, each undergoing Mimaster oscillations, can synchonise the Mixmaster
oscillations of each. The effective coupling will always grow stronger on
approach to the singularity at $\tau =-\infty $ and a critical coupling, $%
K_{crit}$, will always be passed so long as the dynamics are not cut off and
replaced by a different quantum cosmological behaviour at and below the
Planck scale -- only a finite number of Mixmaster oscillations will then
occur and $K_{crit}$ will probably not be reached. On moving away from the
singularity, the coupling will decline and synchronisation will eventually
break down. There is an indication that synchronised behaviour will arise in
the dynamics because in general the shear and Weyl curvature components will
oscillate indefinitely on approach to the singularity \cite{ring2}. Consider
the dimensionless shear variables, $\Sigma _{-}$\ and $\Sigma _{+}$,\ in
eqs. (\ref{4})-(\ref{5}) and put 

\begin{equation}
\Sigma _{+}=\rho \cos \theta ,\text{ \ }\Sigma _{-}=\rho \sin \theta .
\label{shear1}
\end{equation}%
Then, eq.(\ref{9}) determines the evolution of the new variable $\rho $,
while eqns. (\ref{4})-(\ref{5}) show that the angular variable, $\theta ,$%
satisfies

\begin{equation}
\frac{d\theta }{d\tau }=\frac{3}{\rho }(S_{+}\sin \theta -S_{-}\cos \theta ),
\label{shear2}
\end{equation}%
that is\emph{,}

\begin{equation}
\frac{d\theta }{d\tau }=\frac{3}{\rho }\sqrt{\ S_{+}^{2}+S_{-}^{2}}\sin
(\theta -\psi ),  \label{shear3}
\end{equation}

where

\begin{equation}
\tan \psi =\frac{S_{-}}{S_{+}}.  \label{shear4}
\end{equation}

A comparison with eq.(\ref{kur}) shows that the interaction between the
phases $\theta $\ and $\psi $\ suggests a Kuramoto system of equations
leading to synchronisation. The behaviour is more complicated than the basic
Kuramoto model because the coupling is now $\tau $-dependent (see ref.\cite%
{tdep})\emph{.}

\section{Discussion}

Our discussion is of a possible toy model for couplings between different
Mixmaster oscillations. It is a possible effect that has not been considered
previously in attempts to model an inhomogeneous BKL scenario by expansions
around the spatially homogeneous Mixmaster model. It is a picture that has
proven to have very wide applicability in the study of interacting
oscillators in ways that are not specific to individual details of the
physics being modelled. We hope that it will stimulate investigations of new
effects in the general cosmological solutions of the Einstein equations.
There are several obvious features that can be made more realistic. The
constants $\omega _{j}$ and $K$ in eq. \ref{kur} can be made time-dependent
or stochastic with external forcing. Studies of various time-dependent
Kuramoto dynamics have been considered in ref. \cite{tdep}. A time delay
might also be introduced to account for gravitational-wave propagations, as
in ref. \cite{delay} and a hamiltonian formulation may be more suited for a
general relativistic application, see \cite{ham}. We have discussed only the
vacuum solution but the situation in models with pressure and density such
that $p<\rho $ will be similar. In the $p=\rho $ case the chaotic
oscillations always die out on approach to the singularity and the
synchronisation will probably never begin, and very soon end if it does. The
Mixmaster oscillations inside a long era where two scale factors oscillate
approximate to periodic sine and cosine oscillations of the scale factors
when the number of oscillations is very large, but becomes doubly periodic
for smaller numbers of oscillations \cite{pok}. This will introduce other
features of the oscillator couplings which will be reported on elsewhere.

In conclusion, we have introduced consideration of a new factor,
synchronisation of spacetime Mixmaster oscillations, that may play a
simplifying role in understanding the nature of the general inhomogeneous
cosmological solution to Einstein's equations.

\textit{Acknowledgements} The author is supported by the Science and
Technology Facilities Council of the UK (STFC) and thanks V. Belinski and S.
Cotsakis for discussions and a referee for important contributions.

\end{document}